   \documentstyle[12pt]{article}
\if@twoside\oddsidemargin25mm
\else\oddsidemargin25mm\evensidemargin25mm\marginparwidth25mm\fi%
\begin{document}
\newcommand{\js}{{j^\star}}
\newcommand{\K}{K\"ahler\ }
\newcommand{\La}{\Lambda}
\newcommand{\Si}{\Sigma}
\newcommand{\im}{{\rm Im\ }}
\def\bfone{\relax{\rm 1\kern-.35em 1}}
\def\inbar{\vrule height1.5ex width.4pt depth0pt}
\def\IC{\relax\,\hbox{$\inbar\kern-.3em{\rm C}$}}
\def\ID{\relax{\rm I\kern-.18em D}}
\def\IF{\relax{\rm I\kern-.18em F}}
\def\IH{\relax{\rm I\kern-.18em H}}
\def\II{\relax{\rm I\kern-.17em I}}
\def\IN{\relax{\rm I\kern-.18em N}}
\def\IP{\relax{\rm I\kern-.18em P}}
\def\IQ{\relax\,\hbox{$\inbar\kern-.3em{\rm Q}$}}
\def\IR{\relax{\rm I\kern-.18em R}}
\def\IG{\relax\,\hbox{$\inbar\kern-.3em{\rm G}$}}
\font\cmss=cmss10 \font\cmsss=cmss10 at 7pt
\def\ZZ{\relax\ifmmode\mathchoice
{\hbox{\cmss Z\kern-.4em Z}}{\hbox{\cmss Z\kern-.4em Z}}
{\lower.9pt\hbox{\cmsss Z\kern-.4em Z}}
{\lower1.2pt\hbox{\cmsss Z\kern-.4em Z}}\else{\cmss Z\kern-.4em
Z}\fi}
\def\a{\alpha} \def\b{\beta} \def\d{\delta}
\def\e{\epsilon} \def\c{\gamma}
\def\G{\Gamma} \def\l{\lambda}
\def\L{\Lambda} \def\s{\sigma}
\def\cA{{\cal A}} \def\cB{{\cal B}}
\def\cC{{\cal C}} \def\cD{{\cal D}}
\def\cF{{\cal F}} \def\cG{{\cal G}}
\def\cH{{\cal H}} \def\cI{{\cal I}}
\def\cJ{{\cal J}} \def\cK{{\cal K}}
\def\cL{{\cal L}} \def\cM{{\cal M}}
\def\cN{{\cal N}} \def\cO{{\cal O}} \def\cU{{\cal U}}
\def\cP{{\cal P}} \def\cQ{{\cal Q}} \def\cS{{\cal S}}
\def\cR{{\cal R}} \def\cV{{\cal V}}\def\cW{{\cal W}}
%
%
%
\def\crr{\crcr\noalign{\vskip {8.3333pt}}}
\def\tilde{\widetilde}
\def\bar{\overline}
\def\us#1{\underline{#1}}
\let\shat=\hat
\def\hat{\widehat}
\def\hyp{\vrule height 2.3pt width 2.5pt depth -1.5pt}
\def\square{\mbox{.08}{.08}}
\def\Coeff#1#2{{#1\over #2}}
\def\Coe#1.#2.{{#1\over #2}}
\def\coeff#1#2{\relax{\textstyle {#1 \over #2}}\displaystyle}
\def\coe#1.#2.{\relax{\textstyle {#1 \over #2}}\displaystyle}
\def\half{{1 \over 2}}
\def\shalf{\relax{\textstyle {1 \over 2}}\displaystyle}
\def\dag#1{#1\!\!\!/\,\,\,}
\def\to{\rightarrow}
\def\notin{\hbox{{$\in$}\kern-.51em\hbox{/}}}
\def\shdot{\!\cdot\!}
\def\ket#1{\,\big|\,#1\,\big>\,}
\def\bra#1{\,\big<\,#1\,\big|\,}
\def\equaltop#1{\mathrel{\mathop=^{#1}}}
\def\Trbel#1{\mathop{{\rm Tr}}_{#1}}
\def\inserteq#1{\noalign{\vskip-.2truecm\hbox{#1\hfil}
\vskip-.2cm}}
\def\attac#1{\Bigl\vert
{\phantom{X}\atop{{\rm\scriptstyle #1}}\phantom{X}}}
\def\exx#1{e^{{\displaystyle #1}}}
\def\del{\partial}
\def\delbar{\bar\partial}
\def\nex#1{$N\!=\!#1$}
\def\dex#1{$d\!=\!#1$}
\def\cex#1{$c\!=\!#1$}
\def\eg{{\it e.g.}} \def\ie{{\it i.e.}}
%
\newcommand{\be}{\begin{equation}}
\newcommand{\ee}{\end{equation}}
\newcommand{\ba}{\begin{eqnarray}}
\newcommand{\ea}{\end{eqnarray}}
\begin{titlepage}
\hskip 5.5cm
\vbox{ 
       \hbox{POLFIS-TH 02/96}
       \hbox{CERN-TH/95-350}  }
\hskip 1.5cm
\vbox{ \hbox{UCLA/96/TEP/3}
       \hbox{NSF-ITP-96-11}
       \hbox{hep-th/xx9603004}}
\vfill
\begin{center}
{ {\LARGE General Matter Coupled N=2 Supergravity}
\vfill
{ L. Andrianopoli$^1$, M. Bertolini$^2$ $^\dagger$, A. Ceresole$^{2,4}$ \\
\vskip 1.5mm
R. D'Auria$^2$, S. Ferrara$^{3-5}$ and P. Fr\'e$^6$} \\
\vfill
}
{\small
$^1$ Dipartimento di Fisica, Universit\'a di Genova, via Dodecaneso 33,
I-16146 Genova, Italy\\
\vspace{6pt}
$^2$ Dipartimento di Fisica, Politecnico di Torino,\\
 Corso Duca degli Abruzzi 24, I-10129 Torino, Italy\\
\vspace{6pt}
$^3$ CERN, CH 1211 Geneva 23, Switzerland\\
\vspace{6pt}
$^4$ Institute for Theoretical Physics, University of California,
 Santa Barbara, CA 93106, USA\\
\vspace{6pt}
$^5$ Department of Physics, University of California, Los Angeles, CA 90024,
USA\\
\vspace{6pt}
$^6$ Dipartimento di Fisica Teorica, Universit\`a di Torino, via
P. Giuria, 1
I-10125 Torino, Italy\\
}
\end{center}
\vfill
\begin{center} {\bf Abstract}
\end{center}
{
\small
The general form of $N\!=\!2$ supergravity  coupled to an arbitrary
number of vector multiplets and hypermultiplets, with a generic
gauging of the scalar manifold isometries is given. This extends
the results already available in the literature in that we use a 
coordinate independent and manifestly symplectic covariant formalism
which allows to cover theories difficult to formulate within superspace
or tensor calculus approach.
We  provide the  complete lagrangian and supersymmetry variations
with all fermionic terms, and the form of the scalar potential for
arbitrary quaternionic manifolds and special geometry, not necessarily
in special coordinates. Our results can be used to explore properties
of theories admitting $N=2$ supergravity as low energy limit.
}

\vspace{2mm} \vfill \hrule width 3.cm
{\footnotesize
\noindent $^\dagger$ Fellow by Ansaldo Ricerche srl, C.so Perrone 24,
I-16152 Genova\\
\noindent
$^*$ Supported in part by DOE grant
DE-FGO3-91ER40662 Task C., EEC Science Program SC1*CT92-0789,
NSF grant no. PHY94-07194, and INFN.}
\end{titlepage}

\section{Introduction}
Impressive results over the last year on non perturbative properties of $N=2$
supersymmetric Yang-Mills theories\cite{SW12,kltold} 
and their extension to string theory\cite{CDF,CDFVP,huto1,Wdy}
through the notion of string-string duality\cite{Duff}, have used the deep
underlying mathematical structure of these theories and its relation to
algebraic geometry.

In the case of $N=2$ vector multiplets, describing the effective
interactions in the Abelian (Coulomb) phase of a spontaneously
broken gauge theory, Seiberg and Witten \cite{SW12}
have shown that positivity of
the metric on the underlying moduli space identifies the geometrical
data of the effective $N=2$ rigid theory with the periods of
a particular torus.
\par
In the coupling to gravity it was conjectured by some of the present
authors \cite{CDF,CDFVP}
and later confirmed by heterotic-Type II duality
\cite{FHSV,kava,cynoi,KLM},  that the very same argument based on positivity
of the vector multiplet kinetic
metric identifies the corresponding geometrical data of the
effective $N=2$ supergravity with the periods of Calabi-Yau
threefolds.
\par
On the other hand, when matter is added, the underlying geometrical
structure is much richer, since  $N=2$  matter hypermultiplets are
associated with quaternionic geometry\cite{bagwit,hklr,gal}, 
and charged hypermultiplets are
naturally associated with the gauging of triholomorphic isometries of
these quaternionic manifolds \cite{DFF}.

\par
It is the aim of this paper to complete the general form of  the
$N=2$ supergravity lagrangian coupled to an arbitrary number of
vector multiplets and hypermultiplets  in presence of a general gauging 
of the isometries of both the vector multiplets and hypermultiplets 
scalar manifolds.
Actually this extends  
results already obtained years ago by some of us \cite{DFF},
that in turn extended previous work by Bagger and Witten  on ungauged
general quaternionic manifolds coupled to $N=2$ supergravity\cite{bagwit},
by de Wit, Lauwers and Van Proeyen on gauged special geometry and gauged
quaternionic manifolds obtained by quaternionic quotient in the
tensor calculus framework \cite{dWLVP},
and by Castellani, D'Auria and Ferrara on covariant formulation of
special geometry for matter coupled supergravity \cite{CaDF}.
\par
This paper firstly provides in a geometrical setting the full lagrangian
with all the fermionic terms and the supersymmetry variations. Secondly,
it uses a coordinate independent and manifestly
symplectic covariant formalism which in particular does not require
the use of a prepotential function $F(X)$.
Whether a prepotential $F(X)$ exists or not depends on the choice
of a symplectic gauge\cite{CDFVP}. Moreover, some physically interesting
cases are precisely instances where $F(X)$ does not exist\cite{CDFVP}. 
\par
Of particular relevance is the fact that we exhibit a scalar potential for
arbitrary quaternionic geometries
and for special geometry not necessarily in special coordiantes. This
allows us to go beyond what is obtainable with the tensor calculus (or
superspace) approach. 
Among many applications, our results  allow the study of general conditions
for spontaneous supersymmetry breaking in a manner analogous to what was done
for  $N=1$ matter coupled supergravity \cite{CFGVP}.
Many examples of supersymmetry breaking studied in the past are then
reproduced in a unified framework.
\par
Recently the power of using simple geometrical formulae for the
scalar potential was exploited while studying the breaking of half
supersymmetries in a particular simple model, using a symplectic basis
where $F(X)$ is not defined\cite{FGP1}. The method has potential
applications in string theory to study  non perturbative phenomena
such as conifold transitions \cite{GMS}, $p$-forms  condensation 
\cite{pols} and  Fayet-Iliopoulos terms \cite{FGP1,APT}.

\par
Although the supersymmetric Lagrangian and the transformation
rules look quite involved, all the couplings, the mass matrices
and the vacuum energy  are completely fixed and
organized in terms of few geometrical data, such as
the choice of a gauge group $G$, and
of a special K\"ahler ${\cal SK}(n_V)$ and of a Quaternionic manifold
$\cQ (n_H)$ describing the self-interactions of the $n_V$ vector
and $n_H$ hypermultiplets respectively,  whose direct product yields
the full scalar manifold of the theory
\be
{\cal M} \equiv  {\cal SK}(n_V) \otimes {\cal Q}(n_H)\ .
\ee
If $G$ is non-abelian, it must be a subgroup of the isometry
group of the scalar manifold ${\cal SK}(n_V)$  with a block diagonal immersion
in the symplectic group $Sp(2n_V +2,\IR ) $ of
 electric-magnetic duality rotations.

An expanded version of this paper, with particular attention to the
geometrical properties of the scalar manifolds,  the rigidly
supersymmetric version and further related issues is given
in \cite{lungo}.

\section{Resum\'e and  Glossary of Special and Quaternionic Geometry}

\setcounter{equation}{0}

Here we collect some useful formulae for special and quaternionic
geometry, following closely the conventions of \cite{CDFp}.
The $n_V$ complex scalar fields $z^i$ of $N=2$ vector multiplets are
coordinates of a special \K manifold, that is  a K\"ahler-Hodge manifold
${\cal SK}(n_V)$ with
the additional constraint on the curvature
\be
R_{ij^\star k l^\star}=g_{i\js}g_{k l^\star }+g_{i l^\star } g_{k\js}
-C_{ikp} C_{\js l^\star p^\star} g^{p p^\star}\ ,
\label{duno}
\ee
where $g_{i\js}=\del_i\del_{\js}K$ is the \K metric, $K$ is the \K
potential and $C_{ikp}$ is a completely
symmetric covariantly holomorphic tensor.
We remind that the Levi-Civita connection one form and the Riemann
tensor are given by
\be
\Gamma^i_j=\Gamma^i_{kj} dz^k\ ,\ \Gamma^i_{kj}=g^{i l^\star} \del_j g_
{k l^\star}\ ,\ R^i_{\ j k^\star l}=\del_{k^\star } \Gamma^i_{jl}\ .
\ee

A K\"ahler-Hodge manifold has the property that there is a $U(1)$ bundle
$\cL$ whose first Chern class coincides with the \K class. This means that
locally the $U(1)$ connection $Q$ can be written as
\be
Q=-\frac{i}{2}(\del_i K dz^i-\del_{i^\star}K d\bar z^{i^\star})\ .
\ee
The covariant derivative of a generic
field $\psi^i$, that under a \K transformation $K\to K+f+\bar f$ transforms
as $\psi^i\to exp[-{1\over2}(pf+\bar p \bar f )]\psi^i$ is given by
\ba
D_i\psi^j &=& \del_i \psi^j+\Gamma^j_{ik}\psi^k +\frac{p}{2}\del_i K \psi^j\ ,
\nonumber\\
D_{i^\star}\psi^j &=& \del_{i^\star} \psi^j +\frac{\bar p}{2}
\del_{i^\star} K \psi^j\ .
\ea

(In the following we always have $\bar p=-p$). Note that $\bar\psi^{j^\star}$
has weight $(-p,-\bar p)$.
Since $C_{ikp}$ is covariantly holomorphic and has weight $p=2$, it satisfies
$D_{q^\star }C_{ikp}=(\del_{q^\star }-\del_{q^\star } K)C_{ikp}=0$.

A more intrinsic and useful definition of a special \K manifold can be given
by constructing a flat $2n_V+2$-dimensional symplectic bundle over the \K-Hodge
manifold whose generic sections (with weight $p=1$)
\be
V = (L^\La,M_\La)\ \ \ \ \ \La=0,\ldots,n_V\ ,
\label{sezio}
\ee
are covariantly holomorphic
\be
D_{i^\star} V = (\del_{i^\star}-\half\del_{i^\star} K) V=0
\label{ddue}
\ee
and satisfy the further condition
\be
i <V,\bar V >=i(\bar L^\La M_\La-\bar M_\La L^\La)=1\ ,
\label{dtre}
\ee
where $<\ ,\ >$ denotes a symplectic inner product with  metric  chosen to be
$\pmatrix{0 &-\bfone\cr\bfone&0}$.

Defining $U_i =D_i V=(f_i^\La,h_{i\La})$, and introducing a symmetric 
three-tensor $ C_{ijk}$ by
\be
D_i U_j=i C_{ijk} g^{kk^\star}\bar U_{k^\star}\ ,
\label{dqua}
\ee
one can show that the symplectic connection
\ba
D_i V &=& U_i\nonumber\\
D_i U_j &=& i C_{ijk} g^{kk^\star}\bar U_{k^\star}\nonumber\\
D_i U_{j^\star} &=& g_{i j^\star } \bar V\nonumber\\
D_{i} \bar V &=& 0
\label{geospec}
\ea
is flat, provided the constraint (\ref{duno}) is verified.
Furthermore, the \K potential can be computed as a symplectic
invariant from eq. (\ref{dtre}).
Indeed, introducing also the holomorphic sections
\ba
\Omega &=& e^{-K/2} V=e^{-K/2} (L^\La, M_\La)=(X^\La, F_\La)\nonumber\\
\del_{i^\star}\Omega &=& 0
\ea
eq. (\ref{dtre})  gives
\be
K=-\ln i<\Omega,\bar \Omega >=-\ln i(\bar X^\La F_\La- X^\La \bar F_\La)\ .
\label{dtrd}
\ee
From eqs. (\ref{geospec}), (\ref{dtre}) we have
\ba
<V,U_i> &=& 0\ \ \to\ \ X^\La \del_i F_\La-\del_i X^\La F_\La=0\ ,\label{prima}
\\
 <\bar V, U_i> &=& 0\ ,\nonumber\\
\im \cN_{\La\Si} L^\La \bar L^\Si &=& -\half
 \ \ \to\ \ K=-\ln-2(\bar X^\La \im
\cN_{\La\Si} X^\Si)\ ,
\ea
where the complex symmetric $(n_V+1)\times (n_V+1)$ matrix $\cN_{\La\Si}$
is defined through the relations
\be
M_\La=\cN_{\La\Si} L^\Si\ \ ,
\ \  h_{i^\star\La}=\cN_{\La\Si} f^\Si_{i^\star}\ .
\label{defi}
\ee
The \K metric and Yukawa couplings
$C_{ijk}$ can be written in a manifestly symplectic invariant form as
\ba
g_{i j^\star} &=& -i <U_i, U_{ j^\star} >=-2 f^\La_i \im \cN_{\La\Si}
f^\Si_{j^\star}\ ,\label{metri}\\
C_{ijk} &=& <D_i U_j, U_k>\ .
\label{dott}
\ea
It is also useful to define
\ba
U^{\La\Si} &\equiv & f^\La_i g^{i j^\star} f_{j^\star}^\Si=-\half
 (\im \cN)^{-1\La\Si}-\bar L^\La L^\Si\  ,
\label{dsei}
\ea
which is the inverse relation of eq. (\ref{metri}).
Under coordinate transformations, the sections $\Omega$ transform as
\be
\tilde\Omega =e^{-f_\cS (z)}\cS\Omega\ ,
\label{ddoc}
\ee
where $\cS =\pmatrix{A&B\cr C & D\cr}$ is an element of $Sp(2n_V+2,\IR)$,
\be
A^TD-C^TB=\bfone\ ,\ \ \ A^TC-C^TA=B^TD-D^TB=0\ ,
\label{ddov}
\ee
and the factor $e^{-f_\cS (z)}$ is a $U(1)$ \K transformation.
We also note that
\be
\tilde\cN (\tilde X,\tilde F)=(C+D\cN (X,F))(A+B\cN (X,F))^{-1}\ ,
\label{dvdu}
\ee
a relation that simply derives from its definition, eq. (\ref{defi}) .
 
Note that under K\"ahler transformations $K\to K+f+\bar f$ and
$\Omega\to\Omega e^{-f}$. Since $X^\La\to X^\La e^{-f}$, this means that we
can regard, at least locally, the $X^\La$ as homogeneous coordinates on
$\cal SK (n_V)$\cite{BDW}, provided the  matrix
\be
e^a_i(z)=\del_i(X^a/X^0)\ \ \ a=1,\ldots,n_V
\label{dqin}
\ee
is invertible\cite{CDFVP}. In this case, we may set
\be
 F_\La=F_\La(X)
\ee
and then  eq. (\ref{prima}) implies the integrability condition
\be
\frac{\del F_\Si}{\del X^\La}-\frac{\del F_\La}{\del X^\Si}=0\ \ \to\ \ 
F_\La=\frac{\del F(X)}{\del X^\La}
\ee
with
\be
X^\Si\del_\Si F=2F\ .
\label{dsed}
\ee
$F(X(z))$ is the prepotential of $N=2$ supergravity vector multiplet
couplings \cite{BDW}
and ``special coordinates'' correspond to a coordinate choice for which
\be
e^a_i=\del_i(X^\La/X^0)=\delta_i^a\ .
\label{ddic}
\ee
This means $X^0=1$, $X^i=z^i$.
Since $F=\half X^\La F_\La$, under symplectic transformations the
prepotential transforms as
\be
\tilde F(\tilde X)=F(X)+X^\La(C^TB)_\La^\Si F_\Si+\half X^\La
(C^TA)_{\La\Si} X^\Si+\half F_\La(D^TB)^{\La\Si} F_\Si\ ,
\label{dven}
\ee
where
\be
\tilde X=(A+B\cF) X\ \ ,\ \ \cF=F_{\La\Si}=
{{\del^2 F}\over{\del X^\La\del X^\Si}}\ .
\label{dvun}
\ee

In terms of the special coordinates $t^a={{X^a}\over{X^0}}$,
one has $F(X^\La)=(X^0)^2 f(t^a)$, and the \K potential and the metric are
expressed by

\ba
K(t,\bar t) &=& - \ln \ i\left[ 2f-2\bar
 f+(\bar t^a-t^a)(f_a+\bar f_a)\right]\nonumber\\
G_{a\bar b} &=& \del_a
\del_{\bar b} K(t,\bar t)\ .
\label{dquat}
\ea
Eq. (\ref{dvun}) shows that the tranformation $X\to \tilde X$ can be
actually singular, thus implying the non existence of the prepotential
$F(X)$, depending on the choice of symplectic gauge\cite{CDFVP}.
On the other hand,
some  physically interesting cases, such as the
$N=2\to N=1$ supersymmetry breaking \cite{FGP1}, are precisely instances
where $F(X)$ does not exist. On the contrary the prepotential $F(X)$ seems
to be a necessary ingredient in the tensor calculus constructions of
$N=2$ theories that for this reason are not completely general.
This happens because tensor calculus uses special coordinates
from the very start.
\par
Next we turn to the hypermultiplet sector of an $N=2$ theory.
$N=2$ hypermultiplets are field representations of $N=2$ supersymmetry
which contain a pair of left-handed fermions and a quadruple of
real scalars.
$N=2$ supergravity requires that the $4n_H$ scalars $q^u$ of $n_H$
hypermultiplets be the coordinates of a quaternionic 
manifold\cite{bagwit}.

Supersymmetry requires the existence of a principal
$SU(2)$--bundle ${\cal SU}$
that plays for hypermultiplets the same role played by the
the line--bundle ${\cal L}$
in the case of vector multiplets.

A quaternionic manifold is a $4n_H$-dimensional real manifold
endowed with a metric $h$:
\begin{equation}
d s^2 = h_{u v} (q) d q^u \otimes d q^v   \quad ; \quad u,v=1,\dots,
4n_H
\label{qmetrica}
\end{equation}
and three complex structures $
J^x,\
(x=1,2,3)$
that satisfy the quaternionic algebra
\begin{equation}
J^x J^y = - \delta^{xy} \, \bfone \,  +  \, \epsilon^{xyz} J^z
\label{quatalgebra}
\end{equation}
and such that the metric is hermitian with respect to the three complex
structures:
\begin{equation}
h \left( J^x \mbox{\bf X}, J^x \mbox{\bf Y} \right )   =
h \left( \mbox{\bf X}, \mbox{\bf Y} \right ) \quad \quad
  (x=1,2,3)
\label{hermit}
\end{equation}
where ${\bf X},{\bf Y}$ are generic tangent vectors.
The triplet of  two-forms $K^x$
\begin{equation}
K^x = K^x_{u v} d q^u \wedge d q^v \ \  ; \ \
K^x_{uv} =   h_{uw} (J^x)^w_v
\label{iperforme}
\end{equation}
that provides the generalization of the concept of K\"ahler form
occurring in  the complex case, is covariantly closed with respect to an
$SU(2)\simeq Sp(2)$ connection $\omega^x$
\begin{equation}
\nabla K^x \equiv d K^x + \epsilon^{x y z} \omega^y \wedge
K^z    \, = \, 0
\label{closkform}
\end{equation}
with  curvature given by
\begin{equation}
\Omega^x \, \equiv \, d \omega^x +
{1\over 2} \epsilon^{x y z} \omega^y \wedge \omega^z=\lambda K^x
\label{su2curv}
\end{equation}
where $\lambda$ is a real parameter related to the scale of the quaternionic
manifold. Supersymmetry, together with appropriate normalizations for the
kinetic terms in the lagrangian fixes it to the value $\lambda=-1$. 
Introducing a physical normalization one can set $\lambda
 = M_{P}^{-2} \hat{\lambda}$,
$ M_P $ being the Planck mass, so that the
limit of rigid supersymmetry $ M_P\to + \infty $ can be
identified with $\lambda \to 0$\cite{lungo}.
In that case eq. (\ref{su2curv}) defines a flat $SU(2)$ connection relevant
to the hyper\K manifolds of rigid
supersymmetry, treated extensively in \cite{lungo}).
\par
The holonomy group of a generic quaternionic manifold is in
$Sp(2)\times Sp(2 n_H)$.
Introducing flat
indices $\{A,B,C= 1,2\}\ \ , \{\alpha ,\beta ,\gamma = 1,.., 2n_H\}$
  that run,
respectively, in the fundamental representations of $SU(2)$ and
$Sp(2n_H,\IR)$, we can introduce a vielbein 1-form
\begin{equation}
{\cal U}^{A\alpha} = {\cal U}^{A\alpha}_u (q) d q^u
\label{quatvielbein}
\end{equation}
such that
\begin{equation}
h_{uv} = {\cal U}^{A\alpha}_u {\cal U}^{B\beta}_v
\IC_{\alpha\beta}\epsilon_{AB}
\label{quatmet}
\end{equation}
where $\IC_{\alpha \beta} = - \IC_{\beta \alpha}$, $\IC^2=-\bfone$,
$\epsilon_{AB} = - \epsilon_{BA}$, $\epsilon^2=-\bfone$ are, respectively,
 the flat $Sp(2n_H)$
and $Sp(2) \sim SU(2)$ invariant metrics.

The vielbein ${\cal U}^{A\alpha}$ satisfies the metric postulate,
{\rm i.e.} it is covariantly closed with respect
to the $SU(2)$-connection $\omega^z$ and to some $Sp(2n_H,\IR)$-Lie Algebra
valued connection $\Delta^{\alpha\beta} = \Delta^{\beta \alpha}$
of the holonomy group:
\be
\nabla {\cal U}^{A\alpha} \equiv  d{\cal U}^{A\alpha}
+{i\over 2} \omega^x (\epsilon \sigma_x\epsilon^{-1})^A_{\phantom{A}B}
\wedge{\cal U}^{B\alpha} + \Delta^{\alpha\beta} \wedge {\cal U}^{A\gamma} 
\IC_{\beta\gamma}
=0
\label{quattorsion}
\ee
\noindent
where $(\sigma^x)_A^{\phantom{A}B}$ are the standard Pauli matrices.
Furthermore, because of the reality of $h$, ${ \cal U}^{A\alpha}$ satisfies
the reality condition:
\begin{equation}
{\cal U}_{A\alpha} \equiv ({\cal U}^{A\alpha})^* = \epsilon_{AB}
\IC_{\alpha\beta} {\cal U}^{B\beta}
\label{quatreality}
\end{equation}
A stronger version of eq. (\ref{quatmet}) is
\begin{eqnarray}
({\cal U}^{A\alpha}_u {\cal U}^{B\beta}_v + {\cal U}^{A\alpha}_v {\cal
 U}^{B\beta}_u)\IC_{\alpha\beta}&=& h_{uv} \epsilon^{AB}\nonumber\\
({\cal U}^{A\alpha}_u {\cal U}^{B\beta}_v + {\cal U}^{A\alpha}_v {\cal
U}^{B\beta}_u) \epsilon_{AB} &=& h_{uv} {1\over {n_H}} \IC^{\alpha
\beta}\ .
\label{piuforte}
\end{eqnarray}
\noindent
We have also the inverse vielbein ${\cal U}^u_{A\alpha}$ defined by the
equation
\begin{equation}
{\cal U}^u_{A\alpha} {\cal U}^{A\alpha}_v = \delta^u_v\ .
\label{2.64}
\end{equation}

Let us consider the Riemann tensor
\be
{\cal R}^{A\alpha\ B\beta}_{uv}=\Omega^x_{uv}{i\over2}(\epsilon^{-1}\sigma_x
)^{AB} \IC^{\alpha\beta}+\IR^{\alpha\beta}_{uv}\epsilon^{AB}
\ee
where $\IR^{\alpha\beta}_{uv}$ is the field strength of the $Sp(2n_H)
$ connection:
\begin{equation}
d \Delta^{\alpha\beta} + \Delta^{\alpha \gamma} \wedge \Delta^{\delta \beta}
\IC_{\gamma \delta} \equiv \IR^{\alpha\beta} = \IR^{\alpha \beta}_{uv}
dq^u \wedge dq^v\ .
\label{2.66}
\end{equation}
The $\Omega^x$ and $\IR^{\alpha\beta}$ curvature satisfy the following
relations
\begin{eqnarray}
\Omega^x_{A\alpha , B\beta}&=&\Omega^x_{uv} \cU^u_{A\alpha} \cU^v_{B\beta}
=-i\lambda \IC_{\alpha\beta} (\sigma^x \epsilon)_{AB}\nonumber\\
\IR^{\alpha\beta}_{uv}&=&\frac{\lambda}{2}\epsilon_{AB}(\cU^{A\alpha}_u
\cU^{B\beta}_v-\cU^{A\alpha}_v \cU^{B\beta}_u)+\cU^{A\gamma}_u \cU^{B\delta}_v
 \epsilon_{AB}
\IC^{\alpha\rho}\IC^{\beta\sigma} \Omega_{\gamma\delta\rho\sigma}\ ,
\end{eqnarray}
where $\Omega_{\gamma\delta\rho\sigma}$ is a completely symmetric tensor.
The previous equations imply that the quaternionic manifold is
 an Einstein space with Ricci tensor given by
\be
\cR_{uv}=\lambda (2+n_H) h_{uv}\ .
\ee
\section{The Gauging}
\setcounter{equation}{0}
The problem of gauging matter coupled $N=2$ supergravity theories
consists in identifying the gauge group $G$ as a subgroup, at most of
dimension $n_V+1$ of the isometries of the product space
\be
\cM={\cal SK}(n_V)\otimes \cQ(n_H)\ .
\ee
Here we shall mainly consider two cases even if more general situations
are possible. The first is when the gauge group $G$ is non abelian, the
second is when it is the abelian group $G=U(1)^{n_V+1}$.
In the first case supersymmetry requires that $G$ be
a subgroup of the isometries of $\cM$, since the scalars (more precisely,
the sections
$L^\La$) must belong to the adjoint representation of $G$. In such case
the hypermultiplet space will generically split into\cite{dere}
\be
n_H=\sum_i n_i R_i+\half\sum_l n^P_l R^P_l
\ee
where $R_i$ and $R^P_l$ are a set of irreducible representations of
$G$ and $R^P_l$ denote pseudoreal representations.

\par In the abelian case, the special manifold is not required to have any
isometry and if the hypermultiplets are charged with respect to the $n_V+1$
$U(1)$'s, then the $\cQ$ manifold should at least have $n_V+1$ abelian
isometries.

\par The gauging proceeds by introducing $n_V+1$ Killing vectors generically
acting on $\cM$
\begin{eqnarray}
z^i & \to & z^i+\epsilon^\La k^i_\La (z)\nonumber\\
q^u &\to & q^u+\epsilon^\La k^u_\La (q)\ .
\end{eqnarray}
The \K and quaternionic structures of the factors of $\cM$ imply that
$k^i_\La, k^u_\La$ can be determined in terms of ``Killing prepotentials''
which generalize the so called ``D-terms'' of $N=1$ supersymmetric gauge
theories.  The analysis of such prepotentials was carried out in ref. 
\cite{DFF}, and we now briefly summarize the main results.

For the $\cal SK$ manifold, the Killing prepotential is a real function
$\cP_\La$ satisfying
\be
k^i_\La=i\ g^{ij^\star} \del_{j^\star} \cP_\La\ ,
\label{guno}
\ee
with inverse formula
\be
i \cP_\La=\half (k^i_\La \del_i K- k^{i^\star}_\La \del_{i^\star} K)=
k^i_\La \del_i K = - k^{i^\star}_\La \del_{i^\star} K\ .
\label{gdue}
\ee
Formulae (\ref{guno}), (\ref{gdue}) are the results of $N=1$ supergravity.
For special geometry we have the stronger constraint that the gauge group,
in the non abelian case, should be imbedded in the
symplectic group $Sp(2 n_V+2, \IR)$, and thus one must have
\be
\cL_\La V\equiv k^i_\La \del_i V+ k^{i^\star}_\La \del_{i^\star} V
= T_\La V + f_\La V
\ee
for some $T_\La \in Sp(2 n_V+2, \IR)$ Lie algebra,
 and  $f_\La(z)$ corresponding to an infinitesimal \K transformation (
$\cL_\La$
is the Lie derivative acting on the symplectic section $V$).
We consider here the case where $f_\La=0$ and $G$ is a subgroup of the
classical isometries of $\cal SK$ which have a diagonal embedding in the
symplectic group. Since $f_\La=0$, this means that $\cL_\La K=0$ and $\cP_\La$
satisfies eq. (\ref{gdue}). Moreover we have
\be
k^i_\La U_i=T_\La V+i \cP_\La V
\ee
and taking the symplectic scalar product with $\bar V$ we get
\be
\cP_\La=-<\bar V,T_{\La} V>=-e^K<\bar\Omega,T_\La \Omega>\ .
\ee
By using the property
\be
T_\La=\pmatrix{f^\Si_{\La\Delta}& 0\cr 0 & -f^\Si_{\La\Delta}\cr}\ ,
\ee
we finally get the explicit expression
\be
\cP_\La=e^K(F_\Delta f^\Delta_{\La\Si} \bar X^\Si+\bar F_\Delta
 f^\Delta_{\La\Si} X^\Si)
\ee
in terms of the holomorphic sections $\Omega=(X^\La,F_\La)$.

\par For quaternionic manifolds, eq. (\ref{guno}) is replaced by
\be
k^u_\La \Omega^x_{uv}=-\nabla_v \cP^x_\La=-(\del_v \cP^x_\La+\epsilon^{xyz}
 \omega^y_v \cP^z_\La)\ ,
\label{gtre}
\ee
where $\cP^x_\La$ is a triplet of real zero-form prepotentials.
Eq. (\ref{gtre}) can be solved for the Killing vectors in terms of
 the prepotentials as follows
\be
k^u_\La=\frac{1}{6\lambda^2} \sum_{x=1}^{3} h^{vw}
(\nabla_v\cP^x\Omega^x_{wt})h^{tu}\ ,
\ee
where $h^{uv}$ is the inverse quaternionic metric. If the gauge
 group $G$ has structure constants $f^\La_{\Si\Delta}$ the Killing
 vectors $k^i_\La,k^u_\La$ satisfy the following relations
\begin{eqnarray}
i g_{i j^\star}(k^i_\La k^{j^\star}_\Si-k^i_\Si k^{j^\star}_\La)=
f^\Gamma_{\La\Si}\cP_\Gamma\label{gcin}\ ,\nonumber\\
K^x_{uv} k^u_\La k^v_\Si-\frac{\lambda}{2}\epsilon^{xyz} \cP^y_\La \cP^z_\Si
=\half f^\Delta_{\La\Si} \cP^x_\Gamma\ .
\label{gsei}
\end{eqnarray}
Eq. (\ref{gcin}) can be derived from the group relations of
the Killing vectors 
\be
[ k_\La,k_\Si]=f^\Delta_{\La\Si} k_{\Delta}
\ee
together with their relation to the prepotential functions.

\par An important observation comes from the possible existence of Fayet-
Iliopoulos terms in $N=2$ supergravity. This corresponds to a constant shift
in the prepotential functions
\begin{eqnarray}
\cP_\La & \to & \cP_\La+\cC_\La\nonumber\\
\cP_\La^x & \to & \cP^x_\La +\xi^x_\La\ .
\end{eqnarray}
It is important to observe that in $N=2$ special geometry the Killing vectors
and prepotentials satisfy the relation
\be
k^i_\La L^\La=k^{i^\star}_\La\bar L^\La = \cP_\La L^\La=\cP_\La \bar L^\La=0
\label{basta}
\ee
This implies that $\cC_\La =0$. However, a similar relation does not exist
for the quaternionic Killing vectors so that a F-I term in that case is
possible, subject to the constraint (\ref{gsei}). This implies that in a pure
abelian theory with only neutral hypermultiplets we can still have
$\xi^x_\La\neq 0$ provided
\be
\epsilon^{xyz} \xi^x_\La \xi^y_\Si=0
\ee
holds. 
Models of this sort, breaking $N=2 \to N=0$ with vanishing
cosmological constant, were constructed in ref. \cite{flatpotn2} and will
be discussed in the last section.

The gauging procedure can now be performed through the following steps
\cite{DWNi}. One first defines gauge covariant differentials
\ba
\nabla z^i &=& dz^i + g A^\La k^i_\La (z)\nonumber\\
\nabla \bar z^{i^\star} &=& d\bar z^{i^\star}+g A^\La k^{i^\star}_\La (\bar z)
\nonumber\\
\nabla q^u &=& d q^u + g A^\La k^u_\La (q)\ .
\ea
Secondly, one gauges the composite connections by modifying them by means
of killing vectors and prepotential functions.
\ba
 \Gamma^{i}_{\phantom{i}j}\equiv \Gamma^i_{\ jk}dz^k 
& \to &{\hat \Gamma}^{i}_{\phantom{i}j} =
 \Gamma^{i}_{\phantom{i}jk}\nabla z^k +
 g\, A^\Lambda\, \partial_j k^i_\Lambda \nonumber\\
Q\equiv -\frac{\rm i}{2}(\del_i K dz^i-\del_{i^\star} K d\bar z^{i^\star})
 &\to &{\hat Q}= -\frac{\rm i}{2}(\del_i K \nabla z^i-\del_{i^\star} 
K \nabla\bar z^{i^\star}) + g\, A^\Lambda\, {\cal
P}^0_\Lambda \nonumber\\
\omega^x \equiv \omega^x_u dq^u &\to &{\hat \omega}^x = \omega^x_u
\nabla q^u + g\, A^\Lambda\, {\cal
P}^x_\Lambda \nonumber\\
\Delta^{\alpha\beta}\equiv \Delta^{\alpha\beta}_u d q^u &\to &
{\hat  \Delta}^{\alpha\beta}=
\Delta^{\alpha\beta}_u \nabla q^u  + g\, A^\Lambda\,
\partial_u k_\Lambda^v \, {\cal U}^{u   \alpha A}
 \, {\cal U}^\beta_{v  A}\ ,
\label{compogauging}
\ea
where for simplicity we have assumed a single coupling constant. If
there are many coupling constants corresponding to various factors of the
gauge group the formulas are obviously modified.
Correspondingly, the gauged curvatures are:
\begin{eqnarray}
 {\hat R}^{i}_{\phantom{i}j} & = & R^i_{\phantom{i}j\ell^\star k}
 \, \nabla {\bar z}^{\ell^\star} \wedge \nabla z^k \, + \,
 g\, F^\Lambda\, \partial_j k^i_\Lambda \nonumber\\
 {\hat {\cK}}\equiv d\hat Q &= & {\rm i}g_{ij^\star} 
\, \nabla {\bar z}^{i} \wedge
 \nabla z^{j^\star} \,  + \, g\, F^\Lambda\, {\cal
P}^0_\Lambda \nonumber\\
{\hat \Omega}^x &=& \Omega^x_{uv} \, \nabla q^u
\wedge \nabla q^v \, +\,  g\, F^\Lambda\, {\cal
P}^x_\Lambda \nonumber\\
{\hat  {\IR}}^{\alpha\beta}&=&
\IR^{\alpha\beta}_{uv} \,\nabla q^u
\wedge \nabla q^v  \,  +\,  g\, A^\Lambda\,
 \partial_u k_\Lambda^v \, {\cal U}^{u \vert  \alpha A}
\, {\cal U}^\beta_{v \vert A}\ .
\label{compogaugcurv}
\end{eqnarray}

\section{The Complete N=2 Supergravity Theory}
\setcounter{equation}{0}
We are finally ready to write the supersymmetric
invariant action and supersymmetry transformation
rules for a completely general $N=2$ supergravity.
\par
Such a theory includes
 \begin{enumerate}
\item{{\it the gravitational multiplet}
\be
(V^a_\mu, \psi^A,\psi_A, A^0)
\ee
described by the vielbein one-form $V^a$, $(a=0,1,2,3)$ (together with
the spin connection one-form $\omega^{ab}$),
the $SU(2)$ doublet of gravitino one-forms $\psi^A , \psi_A$
($A=1,2$ and the upper or lower position of the index denotes
right, respectively left chirality, namely 
$\gamma_5 \psi_A=-\gamma_5 \psi^A=1$), and
the graviphoton one-form $A^0$}
\item{ $n_V$ {\it vector multiplets}
\be
(A^I, \lambda^{iA},\lambda^{i^\star}_A ,z^i)
\ee
containing a gauge boson one-form $A^I$ ($I=1,\dots,n_V$), a
doublet of gauginos (zero-form spinors) $\lambda^{iA}$,
$\lambda^{{i}^\star}_A$ of left and right chirality respectively,
and a complex scalar field  (zero-form)
$z^i$ ($i=,1,\dots,n_V$). The scalar fields $z^i$ can be regarded
as arbitrary coordinates on the special manifold ${\cal SK}$
of complex dimension $n_V$.
 }
\item{ $n_H$ {\it hypermultiplets}
\be
(\zeta_\alpha, \zeta^\alpha, q^u)
\ee
formed by
a doublet of zero-form spinors, that is the hyperinos 
$\zeta_\alpha\,\zeta^\alpha$\ 
($\alpha=1,\dots,2 n_H$ and here the  lower or upper position of the index
 denotes
left, respectively right chirality), and four real scalar fields $q^u$ ($u=1,
\dots,4 n_H$), that can be regarded as arbitrary coordinates of the quaternionic
manifold $ {\cal Q}$, of real dimension $4n_H$.
As already mentioned, any quaternionic manifold
 has a holonomy group:
\begin{equation}
{\cal H}ol \left (  {\cal Q} \right ) \, \subset  \, SU(2) \, \otimes
\, Sp(2 n_H ,\IR )
\label{ololobis}
\end{equation}
and the index $\alpha$ of the hyperinos transforms in the
fundamental representation of $Sp(2 n_H, \IR)$. }
 \end{enumerate}
The definition of curvatures in the gravitational sector is given by:
\begin{eqnarray}
T^a & \equiv & dV^a-\omega^a_{\ b}\wedge V^b- {\rm i} \, \bar\psi_A\wedge
\gamma^a\psi^A\label{torsdef}\nonumber\\
\rho_A &  \equiv & d\psi_A-{1\over 4} \gamma_{ab} \,
\omega^{ab}\wedge\psi_A+
{{\rm i} \over 2} {\hat {Q}}\wedge \psi_A +
{\hat \omega}_A^{~B}\wedge \psi_B
\equiv \nabla \psi_A \label{gravdefdown}\nonumber \\
\rho^A & \equiv & d\psi^A-{1\over 4} \gamma_{ab} \, \omega^{ab}\wedge\psi^A
-{{\rm i} \over 2} {\hat { Q}}\wedge\psi^A
+{\hat \omega}^{A}_{\phantom{A}B} \wedge \psi^B \equiv \nabla \psi^A
\label{gravdefup} \nonumber\\
R^{ab} & \equiv & d\omega^{ab}-\omega^a_{\phantom{a}c}\wedge \omega^{cb}\ ,
\label{riecurv}
\end{eqnarray}
where $\omega_A^{\ B}=\frac{\rm i}{2}\omega^x (\sigma_x)_A^B$ and
$\omega^A_{\ B}=\epsilon^{AC}\epsilon_{DB}\omega_C^{\ D}$.
 In all the
above formulae the pull--back on space--time through the maps
\begin{equation}
\begin{array}{ccccccc}
z^i & : & M_4 \, \longrightarrow \, {\cal SK} &; &
q^u & : & M_4 \, \longrightarrow \, {\cal Q}
\end{array}
\label{pollobecco}
\end{equation}
is obviously understood. In this way the composite connections
become one-forms on space-time.
\par
In the vector multiplet sector the curvatures and covariant
derivatives are:
\begin{eqnarray}
\nabla z^i &\equiv& dz^i \, + \, g \, A^\Lambda \, k_\Lambda^i
(z)\label{zcurv}\nonumber\\
\nabla {\bar z}^{{i}^\star} &\equiv& d{\bar z}^{{i}^\star} \,
+ \, g \, A^\Lambda \, k_\Lambda^{{i}^\star}
(\bar z)\label{zcurvb}\nonumber\\
\nabla\lambda^{iA} &\equiv & d\lambda^{iA}-{1\over 4} \gamma_{ab} \,
\omega^{ab} \lambda^{iA} -{{\rm i} \over 2} {\hat {Q}}\lambda^{iA}+
{\hat \Gamma}^i_{\phantom{i}j}\lambda^{jA}+{\hat \omega}^{A}_{~B} \wedge
\lambda^{iB}
\label{lamcurv}
\nonumber\\
\nabla\lambda^{{i}^\star}_A &\equiv &d\lambda^{
i^\star}_A-{1\over 4} \gamma_{ab} \,
\omega^{ab}\lambda^{{i}^\star}_A+{{\rm i} \over 2}
{\hat{ Q}}\lambda^{{i}^\star}_A+
{\hat \Gamma}^{{i}^\star}_{\phantom{\bar
{\imath}}{{j}^\star}}\lambda^{{j}^\star}_A
+{\hat \omega}_{A}^{~B} \wedge
\lambda^{{{i}^\star}}_B
\label{lamcurvb}
\nonumber\\
F^\Lambda &\equiv & dA^\Lambda \, +\,{1\over 2} \, g\,
f^\Lambda_{\phantom{\Lambda}\Sigma\Gamma}\, A^\Sigma\,
\wedge\, A^\Gamma\, +\,
\bar L^\Lambda \bar\psi_A\wedge\psi_B
\epsilon^{AB}+L^\Lambda\bar\psi^A\wedge \psi^B\epsilon_{AB}
\label{Fcurv}
\end{eqnarray}
 where $L^\Lambda=e^{\cK \over 2} X^\Lambda$ is the first half (electric)
of the symplectic section introduced in equation
(\ref{sezio}). (The second part $M_\Lambda$ of such  symplectic sections
would appear in the magnetic field strengths if we did introduce
them.)
\par
Finally, in the hypermultiplet sector  the covariant derivatives
are:
\begin{eqnarray}
{\cal U}^{A \alpha} & \equiv & {\cal U}^{A \alpha}_v \nabla q^v
\,\equiv\, {\cal U}^{A \alpha}_v \left (d\,q^v\, +\,
g\,A^{\Lambda}\,k^v_{\Lambda}(q)\right)
\label{ucurv}\nonumber\\
\nabla \zeta _{\alpha} & \equiv & d \zeta _{\alpha} \,-\,{1 \over 4}
\omega ^{ab}\,\gamma _{ab}\,\zeta _{\alpha}
-{{\rm i} \over 2} {\hat { Q}}\,\zeta _{\alpha}\,+
{\hat {\Delta}}_{\alpha}^{\phantom{\alpha}{\beta}}\zeta _{\beta}
\label{iperincurv}\nonumber\\
\nabla \zeta^{\alpha} & \equiv & d \zeta^{\alpha} \,-\,{1 \over 4}
\omega ^{ab}\,\gamma _{ab}\,\zeta^{\alpha}
+{{\rm i} \over 2} {\hat { Q}}\,\zeta^{\alpha}\,+
{\hat {\Delta}}^{\alpha}_{\phantom{\alpha}{\beta}}\zeta^{\beta}
\label{iperincurvb}
\ea
with
\begin{equation}
{\hat {\Delta}}_{\alpha}^{\phantom{\alpha}{\beta}}\,\equiv\,
{\hat {\Delta}}^{\gamma \beta}\,\IC_{\gamma \alpha}\ \ ;\ \ 
\, {\hat {\Delta}}^{\alpha}_{\phantom{\alpha}{\beta}}
\,\equiv\,\IC_{\beta \gamma}\,{\hat {\Delta}}^{\alpha \gamma}
\end{equation}
\par
Note that the \K weights of all spinor fields  are given
by the coefficients of ${\rm i}\hat{ Q}$ in the definition of their
curvatures and covariant derivatives.

Our next task is to write down the $N=2$ space-time
lagrangian and the supersymmetry transformation laws of the fields.
The method employed for this construction is based on the geometrical
approach, review in \cite{CaDFb}, and  a more detailed
derivation is given in the appendices A and B of
\cite{lungo}.
Actually, one solves the Bianchi identities in $N=2$ superspace and then 
constructs the rheonomic superspace Lagrangian in such a way that
the superspace curvatures and covariant derivatives
given by the  solution of the Bianchi identities
are reproduced by the variational equations of motion derived from the
lagrangian.
After this procedure is completed the space-time lagrangian is
immediately retrieved by restricting the superspace p-forms to
space-time.
The resulting action can be split in the  following way:
\ba
S &=& \int\sqrt{-g}\,d^4 \, x \left[ \cL_{k}+\cL_{4f}+\cL'_g\right]\ ,
\nonumber\\
\cL_k  &=&\cL_{kin}^{inv}+\cL_{Pauli} \ ,\nonumber\\
\cL_{4f} &=&\cL_{4f}^{inv}+\cL_{4f}^{non\, inv}\ ,\nonumber\\
\cL'_g &=&\cL_{mass}-V(z,\bar z , q)\ ,
\label{lagr}
\ea
where $\cL_{kin}^{inv}$ consists of the true kinetic terms as well as 
Pauli-like terms containing the derivatives of the scalar fields.
The modifications due to the gauging are contained not only in
$\cL'_g$ but also in the gauged covariant derivatives in the rest of
the lagrangian. We collect the various terms of (\ref{lagr}) in the table
below.
\begin{center}
\begin{tabular}{c}
\null\\
\hline
\null \\
{\it N=2 Supergravity lagrangian}\\
\null \\
\hline
\end{tabular}
\end{center}
\vskip 0.1cm
\ba
\cL_{kin}^{inv} &=& -\half R + 
g_{ij^\star}\nabla^\mu z^i \nabla_\mu \bar z^{j^\star}+
h_{uv}\nabla_\mu q^u \nabla^\mu q^v + 
{{\epsilon^{\mu\nu\lambda\sigma}}\over{\sqrt{-g}}}
\left( \bar\Psi^A_\mu\gamma_\sigma \rho_{A\nu\lambda} 
-  \bar\Psi_{A\mu} \gamma_\sigma \rho^A_{\nu\lambda} \right )
\nonumber\\
&-& {{\rm i}\over2}g_{ij^\star} \left(\bar\lambda^{iA}\gamma^\mu
\nabla_\mu\lambda^{j^\star}_A
+\bar\lambda^{j^\star}_A \gamma^\mu \nabla_\mu\lambda^{iA}\right ) 
-{\rm i}\left (\bar\zeta^\alpha\gamma^\mu\nabla_\mu\zeta_\alpha
+\bar\zeta_\alpha\gamma^\mu \nabla_\mu \zeta^\alpha \right) \nonumber \\
&+& {\rm i}\left(
\bar {\cal N}_{\Lambda\Sigma}{\cal F}^{-\Lambda}_{\mu\nu}{\cal F}^{-\Sigma
\mu\nu} - 
{\cal N}_{\Lambda\Sigma} {\cal F}^{+\Lambda}_{\mu\nu}{\cal F}^{+ \Sigma
{\mu\nu}}\right )+ \Big\{ -g_{ij^\star}
 \nabla_\mu \bar z^{j^\star}\bar\Psi^\mu_A\lambda^{i A}\nonumber\\
&-& 2 {\cal U}^{A\alpha}_u \nabla_\mu q^u 
\bar\Psi_A^\mu \zeta _\alpha
+ g_{ij^\star}  \nabla _\mu \bar z^{j^\star}
\bar\lambda^{iA}\gamma^{\mu\nu}\Psi_{A\nu}
+ 2{\cal U}^{\alpha A}_u\nabla_\mu q^u
\bar\zeta_\alpha \gamma^{\mu\nu}
\Psi_{A\nu}+{\rm h.c.}\Big\}\nonumber\\
\ \\
\cL_{Pauli} &=& \Big\{ {\cal F}^{-\Lambda}_{\mu\nu} \left(
\im{\cal N}\right )_{\Lambda\Sigma}
{\lbrack} 4 L^\Sigma  \bar\Psi^{A\mu}
\Psi^{B\nu}\epsilon_{AB}-4{\rm i} 
{\bar f}^\Sigma_{i^\star}\bar\lambda^{i^\star}_A\gamma^\nu
\Psi_B^\mu\epsilon^{AB}+\nonumber\\
&+& \half
\nabla_i f^\Sigma_j
\bar\lambda^{iA} \gamma^{\mu\nu} \lambda^{jB}\epsilon_{AB}- 
L^\Sigma \bar\zeta_\alpha\gamma^{\mu\nu} \zeta_\beta 
\IC^{\alpha\beta}
{\rbrack}+{\rm h.c.}\Big\}\\
\nonumber\\
{\cal L}_{4f}^{inv} & = & {{\rm i}\over 2} \left(
g_{ij^\star}\bar\lambda^{iA}\gamma_\sigma
  \lambda^{j^\star}_B -
2 \delta^A_B \bar\zeta^\alpha\gamma_\sigma \zeta_\alpha\right )
\bar\Psi_{A \mu} \gamma_\lambda \Psi^B_\nu
{\epsilon^{\mu \nu \lambda \sigma} \over \sqrt{-g}}\nonumber \\
&-& \,{1 \over 6}\,
\left ( C_{ijk} \bar {\lambda}^{iA} \gamma^{\mu} \Psi ^B_{\mu} \,
\bar {\lambda}^{jC} \lambda ^{kD}\,
\epsilon _{AC} \epsilon _{BD} +h.c.\right)\nonumber \\
&-& 2 \bar {\Psi}^A_{\mu} \Psi ^B_{\nu} \,\bar {\Psi}_A^{\mu} \Psi _B^{\nu}
+2g_{i{{j}^\star}}\,\bar {\lambda}^{iA} \gamma _{\mu} \Psi ^B_{\nu} \,
\bar {\lambda}^{{i}^\star}_A \gamma^{\mu} \Psi _B^{\nu}   \nonumber\\
 &+&  {1 \over 4}
\left (R_{i{{j}^\star}l{{k}^\star}}\,  + \,
g_{i{{k}^\star}} \, g_{l{{j}^\star}}\,
 -\, {3 \over 2}\,g_{i{{j}^\star}} \, g_{l{{k}^\star}}\right )
\bar {\lambda}^{iA}\lambda^{lB} \bar {\lambda}^{{j}^\star}_A\lambda^
{{k}^\star}_B \,
\nonumber \\
& +& {1 \over 4} \,g_{i{{j}^\star}} \,
\bar {\zeta}^{\alpha} \gamma _{\mu} \zeta _{\alpha}\,
\bar {\lambda}^{iA} \gamma ^{\mu} \lambda^{{j}^\star}_A\,
+ \, {1 \over 2} \, {\cal R}^{\alpha}_{\beta ts}
\, {\cal U}^t_{A \gamma}\,{\cal U}^s_{B \delta}
 \epsilon ^{AB} \, C ^{\delta \eta}
\bar {\zeta}_{\alpha}\,\zeta _{\eta}\,\bar {\zeta}^{\beta}\,
\zeta ^{\gamma} \nonumber \\
& -& \left[{{\rm i} \over 12} \nabla _m \, C_{jkl}
\bar {\lambda}^{jA}\lambda^{mB} \bar {\lambda}^{kC}\lambda^{lD}
\epsilon _{AC} \epsilon _{BD} +\,h.\,c.\right] \nonumber \\
& +& g_{i{{j}^\star}}\,
\bar {\Psi}^A_{\mu} \lambda ^{{j}^\star}_A  \,
\bar {\Psi}_B^{\mu} \lambda ^{i B}\,+\,
2  \bar {\Psi}^A_{\mu} \zeta ^{\alpha} \bar {\Psi}_A^{\mu}
 \zeta _{\alpha}\,+ \left ( \epsilon _{AB}\, \IC_{\alpha  \beta} \,
 \bar {\Psi}^A_{\mu} \zeta ^{\alpha} \,
  \bar {\Psi}^{B \vert \mu} \zeta ^{\beta} \,+ \,h. c.\right)
\label{4ferminv} 
\\
{\cal L}_{4f}^{non \,inv} & = &   \Big\{
\left (\im {\cal N} \right )_{\Lambda \Sigma}\, \Bigl[
2 L^{\Lambda}\, L^{\Sigma}\left( \bar {\Psi}^A_{\mu}
 \Psi ^B_{\nu}\right)^-
\left( \bar {\Psi}^C_{\mu} \Psi ^D_{\nu}\right)^- \epsilon _{AB} \,
 \epsilon _{CD} \nonumber\\
&-& 8 {\rm i}\,L^{\Lambda}{\bar f}^{\Sigma}_{{i}^\star}\left( 
\bar {\Psi}^A_{\mu} \Psi ^B_{\nu}\right)^-
\left(\bar {\lambda}^{{i}^\star}_A \gamma^{\nu}
 \Psi _B^{\mu}\right)^-\nonumber\\
&-& 2 {\bar f}^{\Lambda}_{{i}^\star}
 {\bar f}^{\Sigma}_{{j}^\star}
\left(\bar {\lambda}^{{i}^\star}_A \gamma^{\nu} \Psi _B^{\mu}\right)^-
\left(\bar {\lambda}^{{j}^\star}_C \gamma _{\nu} \Psi _{D \vert \mu}\right)^-
\epsilon^{AB}\,\epsilon^{CD} \nonumber\\
&+&{{\rm i} \over 2}
L^{\Lambda}{\bar f}^{\Sigma}_{{\ell}^\star} \,g^{k {\ell}^\star}\,C_{ijk}
\left( \bar {\Psi}^A_{\mu} \Psi ^B_{\nu}\right)^-
\bar {\lambda}^{iC} \gamma^{\mu \nu} \lambda ^{jD}\,
\epsilon _{AB} \,\epsilon _{CD}
\nonumber \\
& +& {\bar f}^{\Lambda}_{ m^\star}{\bar f}^{\Sigma}_{{\ell}^\star}
 \,g^{k {\ell}^\star}\,C_{ijk}
\left(\bar {\lambda}^{m^\star}_A \gamma _{\nu} \Psi _{B \mu}\right)^-
\bar {\lambda}^{iA} \gamma^{\mu \nu} \lambda ^{jB}\nonumber\\
&-& L^{\Lambda} L^{\Sigma}\left( \bar {\Psi}^A_{\mu} \Psi ^B_{\nu}\right)^-
\bar {\zeta}_{\alpha} \gamma^{\mu \nu} \zeta _{\beta} \,
\epsilon _{AB}\, \IC^{\alpha  \beta} \nonumber\\
 & + & {\rm i} L^{\Lambda}{\bar f}^{\Sigma}_{{i}^\star}
\left(\bar {\lambda}^{{i}^\star}_A \gamma^{\nu} \Psi _B^{\mu}\right)^-
\bar {\zeta}_{\alpha} \gamma_{\mu \nu} \zeta _{\beta} \,
\epsilon^{AB} \IC^{\alpha  \beta}
\nonumber \\
& -& {1 \over 32}\,
C_{ijk}\,C_{lmn} g^{k{\bar r}} \, g^{n{\bar s}} \,
{\bar f}^{\Lambda}_{\bar r} \, {\bar f}^{ \Sigma}_{\bar s} \,
\bar {\lambda}^{iA} \, \gamma _{\mu \nu} \,\lambda^{jB} \,
\bar {\lambda}^{kC} \, \gamma^{\mu \nu}\,\lambda^{lD}
\, \epsilon _{AB} \epsilon _{CD}
\nonumber \\
& - & {1 \over 8}\,
L^{\Lambda} \nabla _i f^{\Sigma}_j
\bar {\zeta}_{\alpha} \gamma _{\mu \nu} \zeta _{\beta}\,
\bar {\lambda}^{iA} \gamma ^{\mu \nu} \lambda^{jB}\,
\,\epsilon _{AB} \,\IC^{\alpha  \beta}
\nonumber \\
&+& {1\over 8}\,
L^{\Lambda} \,L^{\Sigma}
\bar {\zeta}_{\alpha} \gamma _{\mu \nu} \zeta _{\beta}\,
\bar {\zeta}_{\gamma} \gamma^{\mu \nu} \zeta _{\delta}\,
\IC^{\alpha  \beta}\,\IC^{\gamma \delta} \Bigr]   +{\rm h.c.}\Big\}
\label{4fermnoninv}\\
\nonumber\\
{\cal L}_{mass}
&=&  g\Big[ 2 S_{AB} \bar\Psi^A_\mu \gamma^{\mu\nu}\Psi^B_\nu +
{\rm i} g_{ij^\star} W^{iAB} \bar\lambda^{j^\star}_A\gamma_\mu \Psi_B^\mu+
 2{\rm i} N^A_\alpha\bar\zeta^\alpha\gamma_\mu \Psi_A^\mu\nonumber\\
&+&
{\cal M}^{\alpha\beta}{\bar \zeta}_\alpha
\zeta_\beta +{\cal M}^{\alpha}_{\phantom{\alpha}iB}
{\bar\zeta}_\alpha \lambda^{iB} + {\cal M}_{iA\ell B}
{\bar \lambda}^{iA} \lambda^{\ell
B}  \Big] + \mbox{h.c.} \\
{\rm V}\bigl ( z, {\bar z}, q \bigr )
&=& g^2 \Bigl[\left(g_{ij^\star} k^i_\Lambda k^{j^\star}_\Sigma+4 h_{uv}
k^u_\Lambda k^v_\Sigma\right) \bar L^\Lambda L^\Sigma
+ g^{ij^\star} f^\Lambda_i f^\Sigma_{j^\star}
{\cal P}^x_\Lambda{\cal P}^x_\Sigma
-3\bar L^\Lambda L^\Sigma{\cal P}^x_\Lambda
{\cal P}^x_\Sigma\Bigr]\ .\\
\nonumber
\end{eqnarray}
where  $\cF^{\pm\La}_{\mu\nu}=\half (\cF^\La_{\mu\nu}\pm
{{\rm i}\over2}\epsilon^{\mu\nu\rho\sigma}\cF_{\rho\sigma}^\La )$ and
$(...)^-$ denotes the self dual part of
the fermion bilinears. The mass--matrices are given by:
\begin{eqnarray}
S_{AB}&=&{{\rm i}\over2} (\sigma_x)_A^{\phantom{A}C} \epsilon_{BC}
{\cal P}^x_{\Lambda}L^\Lambda \nonumber\\
W^{iAB}&=&\epsilon^{AB}\,k_{\Lambda}^i \bar L^\Lambda\,+\,
{\rm i}(\sigma_x)_{C}^{\phantom{C}B} \epsilon^{CA} {\cal P}^x_{\Lambda}
g^{ij^\star} {\bar f}_{j^\star}^{\Lambda}\nonumber\\
N^A_{\alpha}&=& 2 \,{\cal U}_{\alpha u}^A \,k^u_{\Lambda}\,
\bar L^{\Lambda}\nonumber\\
{\cal M}^{\alpha\beta}  &=&-
\, {\cal U}^{\alpha A}_u \, {\cal U}^{\beta B}_v \, \varepsilon_{AB}
\, \nabla^{[u}   k^{v]}_{\Lambda}  \, L^{\Lambda} \nonumber\\
{\cal M}^{\alpha }_{\phantom{\alpha} iB} &=&-
4 \, {\cal U}^{\alpha}_{B  u} \, k^u_{\Lambda} \,
 f^{\Lambda}_i \nonumber\\
{\cal M}_{iA\vert \ell B} &=& \,{1 \over 3}  \,
\Bigl ( \varepsilon_{AB}\,  g_{ij^\star}   k^{j^\star}_ \Lambda  
 f_\ell^\Lambda +
{\rm i}\bigl ( \sigma_x \epsilon^{-1} \bigr )_{AB} \, {\cal P}^x_ \Lambda
\, \nabla_\ell f^\Lambda _i \Bigr )
\label{pesamatrice}
\end{eqnarray}
The coupling constant $g$ in $\cL'_{g}$ is just a symbolic notation to 
remind that these terms are entirely
due to the gauging and vanish in the ungauged theory, where also all
 gauged covariant
derivatives reduce to ordinary ones. Note that in general
there is not a single coupling constant, but rather there are
as many independent coupling constants as the number of factors in
the gauge group. The normalization of the kinetic term for the quaternions
depends on the scale $\lambda$ of the quaternionic manifold, appearing in
eq. (\ref{su2curv}), for which we have chosen the value $\lambda=-1$.

\par Furthermore, using the geometric approach, the form of the supersymmetry
transformation laws is also easily deduced from the solution of the
Bianchi identities in superspace \cite{lungo}by interpreting the
 supersymmetry variations as superspace lie derivatives. One gets
\begin{center}
\begin{tabular}{c}
\null\\
\hline
\null \\
{\it Supergravity transformation rules of the Fermi  fields}\\
\null \\
\hline
\end{tabular}
\end{center}
\vskip 0.2cm
\begin{eqnarray}
\delta\,\Psi _{A \mu} &=& {\cal D}_{\mu}\,\epsilon _A\,
 -{1 \over 4}
 \left(\partial _i\,{K} \bar {\lambda}^{iB}\epsilon _B\,-\,
 \partial _{{i}^\star}\,{ K}
 \bar {\lambda}^{{i}^\star}_B \epsilon^B \right)\Psi _{A  \mu}\nonumber\\
&&-{\omega}_{A  v}^{\phantom{A}B}\,
{\cal U}_{C \alpha}^v\,
\left(\epsilon^{CD}\,\IC^{\alpha  \beta}\,\bar
{\zeta}_{\beta}\,\epsilon _D\,+\,
\,\bar {\zeta}^{\alpha}\,\epsilon^C \right)\Psi _{B  \mu}\nonumber\\
&& +\,\left ( A_{A}^{\phantom{A} {\nu}B}
\eta_{\mu \nu}+A_{A}^{\prime \phantom{A}
  {\nu} B}\gamma_{\mu \nu} \right ) \epsilon _B \,
\nonumber\\
&& + \left [ {\rm i} \, g \,S_{AB}\eta _{\mu \nu}+
\epsilon_{AB}( T^-_{\mu \nu}\, + \,U^+_{\mu \nu} )
\right ] \gamma^{\nu}\epsilon^B
 \label{trasfgrav}\\
\delta\,\lambda^{iA}&=&
{1 \over 4} \left(\partial _j\,{K} \bar {\lambda}^{jB}\epsilon _B\,-\,
\partial _{{j}^\star}\, {K}
 \bar {\lambda}^{{j}^\star}_B \epsilon^B \right)\lambda^{iA}\nonumber\\
&&-{\omega}^{A}_{\phantom{A}B  v}\,
{\cal U}_{C \alpha}^v\,
\left(\epsilon^{CD}\,\IC^{\alpha  \beta}\,\bar
{\zeta}_{\beta}\,\epsilon _D\,+\,
\,\bar {\zeta}^{\alpha}\,\epsilon^C \right)\lambda^{iB}\nonumber\\
&&-\,\Gamma^i_{\phantom{i}{jk}}
\bar {\lambda}^{kB}\epsilon _B\,\lambda^{jA}
+ {\rm i}\, \left (\nabla _ {\mu}\, z^i -\bar {\lambda}^{iA}\psi
_{A \mu}\right)
\gamma^{\mu} \epsilon^A \nonumber\\
&&+G^{-i}_{\mu \nu} \gamma^{\mu \nu} \epsilon _B \epsilon^{AB}\,+\,
D^{iAB}\epsilon _B
\label{gaugintrasfm}\\
\delta\,\zeta _{\alpha}&=&-\Delta _{\alpha  v}^{\phantom{\alpha}\beta}\,
{\cal U}_{\gamma A}^v\,
\left(\epsilon^{AB}\,\IC^{\gamma \delta}\,\bar
{\zeta}_{\delta}\,\epsilon _B\,+\,
\,\bar {\zeta}^{\gamma}\,\epsilon^A \right)\zeta _{\beta}\nonumber\\
&&+ {1 \over 4} \left(\partial _i\,{ K} \bar {\lambda}^{iB}\epsilon _B\,-\,
 \partial _{{i}^\star}\,
{ K} \bar {\lambda}^{{i}^\star}_B
 \epsilon^B \right)\zeta _{\alpha}\nonumber\\
 &&+\, {\rm i}\,\left(
{\cal U}^{B \beta}_{u}\, \nabla _{\mu}\,q^u\,
-\epsilon^{BC}\,\IC^{\beta\gamma}\,\bar
{\zeta}_{\gamma}\,\psi _C\,-\,
\,\bar {\zeta}^{\beta}\,\psi^B
\right)\,\gamma^{\mu} \epsilon^A
\epsilon _{AB}\,\IC_{\alpha  \beta}
\,+\,g\,N_{\alpha}^A\,\epsilon _A \label{iperintrasf}
\end{eqnarray}
\begin{center}
\begin{tabular}{c}
\null\\
\hline
\null \\
{\it Supergravity transformation rules of the Bose  fields}\\
\null \\
\hline
\end{tabular}
\end{center}
\vskip 0.2cm
\ba
\delta\,V^a_{\mu}&=& -{\rm i}\,\bar {\Psi}_{A 
\mu}\,\gamma^a\,\epsilon^A -{\rm i}\,\bar {\Psi}^A _
\mu\,\gamma^a\,\epsilon_A\\
\delta\,A^\Lambda _{\mu}&=& 
2 \bar L^\Lambda \bar \psi _{A\mu} \epsilon _B
\epsilon^{AB}\,+\,2L^\Lambda\bar\psi^A_{\mu}\epsilon^B \epsilon
_{AB}\nonumber\\
&+&\left({\rm i} \,f^{\Lambda}_i \,\bar {\lambda}^{iA}
\gamma _{\mu} \epsilon^B \,\epsilon _{AB} +{\rm i} \,
{\bar f}^{\Lambda}_{{i}^\star} \,\bar\lambda^{{i}^\star}_A
\gamma _{\mu} \epsilon_B \,\epsilon^{AB}\right)\label{gaugtrasf}\\
\delta\,z^i &=& \bar{\lambda}^{iA}\epsilon _A \label{ztrasf}\\
\delta\,z^{{i}^\star}&=& \bar{\lambda}^{{i}^\star}_A \epsilon^A
\label{ztrasfb}\\
  \delta\,q^u &=& {\cal U}^u_{\alpha A} \left(\bar {\zeta}^{\alpha}
  \epsilon^A + \IC^{\alpha  \beta}\epsilon^{AB}\bar {\zeta}_{\beta}
  \epsilon _B \right)
 \ea
where we have:
\begin{center}
\begin{tabular}{c}
\null\\
\hline
\null \\
{\it Supergravity values of the auxiliary  fields}\\
\null \\
\hline
\end{tabular}
\end{center}
\vskip 0.2cm
\begin{eqnarray}
A_{A}^{\phantom{A}  \mu B}
&=&-{{\rm i} \over 4}\, g_{{{k}^\star}\ell}\,
\left(\bar {\lambda}^{{k}^\star}_A \gamma^{\mu} \lambda^{\ell B}\,
-\,\delta^B_A\,
\bar\lambda^{{k}^\star}_C \gamma^{\mu} \lambda^{\ell C}\right)\label{Adef}\\
A_{A}^{\prime \phantom{A} \mu B}
&=&{{\rm i} \over 4}\, g_{{{k}^\star}\ell}\,\left(\bar{\lambda}^{{k}^\star}_A
 \gamma^{\mu}
\lambda^{\ell B}-{1\over 2}\, \delta^B_A\, \bar\lambda^{{k}^\star}_C
\gamma^{\mu} \lambda^{C\ell}\right) \, -\,  {{\rm i} \over 4}\,
 \delta _A^B \,\bar \zeta _{\alpha}
\gamma^{\mu} \zeta^{\alpha}\label{A'def}
\end{eqnarray}
\begin{eqnarray}
T^-_{\mu\nu} &=& 2{\rm i}
\left (\im {\cal N}\right)_{\Lambda\Sigma} L^{\Sigma}
\left ({\tilde{F}}_{\mu\nu}^{\Lambda -} +{1\over 8} \nabla_i 
\,f^{\Lambda} _j \,
\bar \lambda^{i A} \gamma_{\mu\nu} \, \lambda^{jB} \,\epsilon_{AB}
-{1\over 4} \, \IC^{\alpha  \beta}\,{\bar\zeta}_{\alpha}\gamma _{\mu\nu} \,
\zeta _{\beta}\, L^{\Lambda}
\right )\label{T-def}\\
T^+_{\mu\nu} &=& 2 {\rm i}
\left (\im {\cal N}\right)_{\Lambda\Sigma} {\bar L}^{\Sigma}
\left({\tilde{F}}_{\mu\nu}^{\Lambda +} +{1\over 8} \nabla_{{i}^\star} \,\bar
f^\Lambda_{{j}^\star}\,
\bar \lambda^{{i}^\star}_A \gamma _{\mu\nu} \, \lambda^{{j}^\star}_B
 \epsilon^{AB}
-{1 \over 4}\, \IC_{\alpha  \beta}\,{\bar\zeta}^{\alpha}\gamma _{\mu\nu} \,
\zeta ^{\beta}\, {\bar L}^{\Lambda}
\right)
\label{T+def}
\end{eqnarray}
\begin{eqnarray}
U^-_{\mu\nu} &=& -{{\rm i} \over 4} \, \IC^{\alpha  \beta}\,
{\bar\zeta}_{\alpha}\gamma _{\mu\nu} \,
\zeta _{\beta}\label{U-def}\\
U^+_{\mu\nu} &=& -{{\rm i} \over 4} \,
 \IC_{\alpha  \beta}\,{\bar\zeta}^{\alpha}\gamma _{\mu\nu} \,
\zeta^{\beta}\label{U+def}
\end{eqnarray}
\begin{eqnarray}
G^{i-}_{\mu\nu} &=& - g^{i{{j}^\star}}
 \bar f^\Gamma_{{j}^\star}
\left (\im  {\cal N}\right)_{\Gamma\Lambda}
\Bigl ( {\tilde {F}}^{\Lambda -}_{\mu\nu} + {1\over 8}
\nabla_{k}  f^{\Lambda}_{\ell} \bar \lambda^{kA}
\gamma_{\mu\nu} \, \lambda^{\ell B} \epsilon_{AB}
  \nonumber\\
&\null& \qquad \qquad  -
{1\over 4} \, \IC^{\alpha  \beta}\,{\bar\zeta}_{\alpha}\gamma _{\mu\nu}
\, \zeta _{\beta}\, L^{\Lambda}
\Bigr )\label{G-def}\\
G^{{{i}^\star}+}_{\mu\nu} &=& - g^{{{i}^\star}j} f^{\Gamma}_j
\left (\im {\cal N}\right)_{\Gamma\Lambda}
\Bigl ( {\tilde{F}}^{\Lambda +}_{\mu\nu} +{1\over 8}
\nabla _{{k}^\star} \bar f^{\Lambda}_{{\ell}^\star} \bar \lambda^{{k}^\star}_A
\gamma _{\mu\nu} \, \lambda^{{\ell}^\star}_B \epsilon^{AB}\nonumber\\
&\null& \qquad \qquad - {1 \over 4}\,
\IC_{\alpha  \beta}\,{\bar\zeta}^{\alpha}\gamma _{\mu\nu} \,
\zeta ^{\beta}\, {\bar L}^{\Lambda}
\Bigr )\label{G+def}
\end{eqnarray}
\begin{eqnarray}
D^{iAB} &=& {{\rm i} \over 2}g^{i{{j}^\star}}
 C_{{{j}^\star}{{k}^\star}{{\ell}^\star}} \bar\lambda^{{k}^\star}_C
\lambda^{{\ell}^\star}_D
\epsilon^{AC} \epsilon^{BD}\,+\,W^{iAB}\label{3.21a}
\end{eqnarray}
In eqs. (\ref {T-def}), (\ref{T+def}), (\ref{G-def}), (\ref{G+def})
 we have denoted by ${\tilde{F}}_{\mu\nu}$
the supercovariant field strength defined by:
\begin{equation}
\tilde{F}^\Lambda_{\mu\nu} = \cF^\Lambda_{\mu\nu}\,
+\,L^{\Lambda}\bar\psi^A_\mu\psi^B_{\nu}
\,\epsilon_{AB} \,+\bar L^{\Lambda}\bar{\psi}_{A \mu} \psi_{B \nu}
\epsilon^{AB}\,
-{\rm i} \,f^{\Lambda}_i \,\bar {\lambda}^{iA} \gamma_{[\nu}
 \psi^B_{\mu]}\,\epsilon _{AB}\,
-{\rm i} \,{\bar f}^{\Lambda}_{{i}^\star} \,\bar\lambda^{i^\star}_A
\gamma_{[\nu} \psi _{B \mu]} \,\epsilon^{AB}\ .
\end{equation}
Let us make some observation about the structure of the Lagrangian
and of the transformation laws.
\par
i) We note that all the terms of the lagrangian are given in terms of
purely geometric objects pertaining to the special and quaternionic
geometries. Furthermore the Lagrangian does not rely on the existence
of a prepotential function $F=F\left(X\right)$ and it is valid for
any choice of the quaternionic manifold.
\par
ii) The lagrangian is not invariant under symplectic duality
transformations. However, in absence of gauging ($g=0$), if we
restrict the lagrangian to configurations where the vectors are
on shell, it becomes symplectic invariant\cite{fsz,CDFVP}. This allows us
to fix the terms appearing in ${\cal L}^{non\,inv}_{4f}$ in
a way independent from supersymmetry arguments.
\par
iii) We note that the field strengths
${\cal F}^{\Lambda\,-}_{\mu\nu}$ originally introduced in the Lagrangian 
are the free gauge field
strengths. The interacting field strengths which are supersymmetry
eigenstates are defined as the objects appearing in the
transformation laws of the gravitinos and gauginos
fields, respectively, namely the bosonic part of $T^-_{\mu\nu}$ and
$G^{-\,i}_{\mu\nu}$ defined in eq.s (\ref{T-def}), (\ref{G-def}).

\section{Comments on the scalar potential}
A general Ward identity\cite{ward} of $N$-extended supergravity establishes the
following formulae for the scalar potential $V(\phi)$ of the theory
(in  appropriate normalizations for the generic fermionic shifts $\delta
\chi^a$)
\be
Z_{ab} \delta_A \chi^a \delta^B\bar \chi^b-3 \cM_{AC}\bar\cM^{CB}=
\delta^A_{\ B}  V(\phi)\ \ \ A,B=1,\ldots,N
\ee
where $\delta_A\chi^a$ is the extra contribution, due to the gauging,
 to the spin $\half$
supersymmetry variations of the scalar vev's, $Z_{ab}$ is the (scalar
 dependent) kinetic term normalization and $\cM_{AC}$ is the (scalar
dependent) gravitino mass matrix. Since in the case at hand ($N=2$) all
terms in question are expressed in terms of Killing vectors and
prepotentials, contracted with the symplectic sections, we will be
able to derive a completely geometrical formula for $V(z, \bar z,q)$.
The relevant terms in the fermionic transformation rules are
\begin{eqnarray}
\delta \psi_{A\mu}&=&ig S_{AB} \gamma_\mu \epsilon^B\ ,\nonumber\\
\delta\lambda^{iA}&=& g W^{iAB}\epsilon_B\ ,\nonumber\\
\delta\zeta _ \alpha &=& g N^A_\alpha \epsilon_A\ .
\end{eqnarray}
In our normalization the previous Ward identity gives
\begin{equation}
V=(g_{ij^\star}k^i_\La k^{j^\star}_\Si +4 h_{uv} k^u_\La k^v_\Si) 
\bar L^\La L^\Si+(U^{\La\Si}-3 \bar L^\La L^\Si )
\cP^x_\La \cP^x_\Si\ .
\label{ssette}
\end{equation}
with $U^{\La\Si}$ is defined in (\ref{dsei}).
Above, the first two terms  
are related to the gauging of isometries of ${\cal SK}\otimes
\cQ$. For an abelian group, the first term  is
absent. The negative term is  the gravitino mass
contribution, while the one in $U^{\La\Si}$ is the gaugino shift
contribution due to the quaternionic prepotential.

Eq. (\ref{ssette}) can be rewritten in a suggestive form as
\be
V=(k_\La ,k_\Si )\bar L^\La L^\Si+(U^{\La\Si}-3 \bar L^\La L^\Si )
(\cP_\La^x \cP_\Si^x-\cP_\La\cP_\Si)\ ,
\label{sotto}
\ee
where
 \begin{equation}
(k_\La , k_\Si )=\pmatrix{k^i_\La ,&k^{i^\star}_\La ,&k^u_\La}
\pmatrix{0&g_{ij^\star}&0\cr g_{i^\star j}&0&0\cr0&0&2h_{uv}}
\pmatrix{k^j_\Si\cr k^{j^\star}_\Si\cr k^v_\Si}
\label{snove}
\end{equation}
is the scalar product of the Killing vector and we have used eqs.
(\ref{guno}),(\ref{basta}) .  $\cP^x_\La$ are
the quaternionic (triplet) prepotentials and $U^{\La\Si}, L^\La $
are special geometry data.

In a theory with only abelian vectors, the potential may still be 
non-zero due to Fayet-Iliopoulos terms:
\be
\cP^x_\La=\xi_\La^x\ ({\rm constant});\ \ \
\epsilon^{xyz}\xi^y_\La \xi^z_\Si=0 \ .
\ee
In this case
\be
V(z,\bar z)= (U^{\La\Si}-3 \bar L^\La L^\Si)\xi^x_\La \xi^x_\Si\ .
\ee
Examples with $V(z,\bar z)=0$ but non-vanishing gravitino mass (with
$N=2$ supersymmetry broken to $N=0$) were given in \cite{flatpotn2},
 then generalizing to $N=2$ the no scale models of $N=1$ supergravity
 \cite{flatpotn1}.
These models were obtained by taking a $\xi^x_\Lambda=(\xi_0,0,0)$ .
In this case the expression

\be
V= U^{00}-3 \bar L^0 L^0
\ee
reduces to
\be
V=(\partial _i K g^{i j^\star} \partial _{j^\star} K -3)e^K
\ee
which is the $N=1$ supergravity potential, with solution ( $V=0$) the cubic
 holomorphic prepotential

\be
F(X)=d_{ABC}\frac{X^A X^B X^C}{X^0}\ \ \ \ A=1,\ldots,n \ .
\ee

Another solution is obtained by taking the $\frac{SU(1,1)}{U(1)}
\otimes
\frac{SO(2,n)}{SO(n)}$ coset in the $SO(2,n)$ symmetric parametrization of
the symplectic sections ($X^\Lambda,F_\Lambda=\eta_{\La\Si} S X^\Si\ ;
\ X^\La X^\Si \eta_{\La\Si}=0\ ,\eta_{\La\Si}=(1,1,-1,\ldots,-1)$)
where a prepotential $F$ does not exist. In this case
\be
U^{\La\Si}-3 \bar L^\La L^\Si=-\frac{1}{i(S-\bar S)}\eta_{\La\Si}
\label{genov}
\ee
where we have used the fact that
\be
\cN_{\La\Si}=(S-\bar S) (\Phi_\La \bar\Phi_\Si+\bar\Phi_\La \Phi_\Si)
+\bar S\eta_{\La\Si}\ \ \ ,\ \Phi^\La=\frac{X^\La}{(X^\La \bar X_\La)^{1/2}}
\ .
\ee
The identity (\ref{genov}) allows one to prove that the tree level potential
of an arbitrary heterotic string compactification (including orbifolds with
twisted hypermultiplets) is semi-positive definite provided we don't gauge the
graviphoton and the gravidilaton vectors ($\rm i.e.$ $\cP^x_\La=0$ for
$\La=0,1$, $\cP^x_\La\neq 0$ for $\La=2,\ldots,n_V$). On the other hand, it
also proves that tree level supergravity breaking may only occurr if
$\cP^x_\La\neq 0$ for $\La=0,1$. This instance is related to models with
Scherk-Schwarz mechanism studied in the literature \cite{scsc,ant}. 

A vanishing potential can be obtained if $\xi^x_\La=(\xi_\La,0,0)$ with
\be
\xi_\La \xi_\Si \eta^{\La\Si}=0\ .
\ee
In this case we may also consider the gauge group to be $U(1)^{p+2}\otimes
 G(n_V-p)$ and introduce $\xi _ \Lambda \,=\, \left(\xi_0,\ldots,\xi_{p+1},0,
\ldots,0 \right)$
 such that
$\xi_\La \xi_\Si \eta^{\La\Si}=0$ where $\eta^{\La\Si}$ is the $SO(2,p)$
Lorentzian metric. The potential is now:
\be
V=k^i_\La g_{i j^\star} k^{j^\star}_\Si \bar L^\La L^\Si\ \ ;\ \
(U^{\La\Si}-3\bar L^\La L^\Si) \cP^{x}_\La \cP^{x}_\Si=0
\ee
where $k^i_\La L^{\Lambda}=0$ for $\La \leq p+1$. The gravitino have
equal mass
\be \mid m_{3/2} \mid \simeq e^{K/2} \mid \xi_\La X^\La \mid
\ee
with $\xi_\La \xi_\Si \eta^{\La\Si}=0$ , $\La=0, \ldots, p+1$.

It is amusing to note that the gravitino mass, as a function of the
$O(2,p)/O(2)\otimes O(p)$ moduli and of the F-I terms, just coincides with
 the central charge formula for the level $N_L=1$ in heterotic string
(H-monopoles), if the F-I terms are identified with the $O(2,p)$ lattice
electric charges.

Note that, because of the special form of the gauged $\hat Q$,
 $\hat \omega^x$, we see that whenever $\cP_\La\neq 0$ the gravitino is
charged with respect to the $U(1)$ factor and whenever $\cP^x_\La \neq 0$
 the gravitino is charged with respect to the $SU(2)$ factor of the
 $U(1)\otimes SU(2)$ automorphism group 
of the supersymmetry algebra. In the case of
$U(1)^p$ gauge fields with non-vanishing F-I terms $\xi^x_p=(0,0,\xi_p)$ the
 gauge field $A^\La_\mu \xi_\La=A_\mu$ gauge a $U(1)$ subgroup 
of $SU(2)_L$ susy algebra.

Models with breaking of $N=2$ to $N=1$ \cite{FGP1}
necessarily require $k^u_\La$ not to be zero. The
minimal model where this happens with $V=0$ is the one based on
\be
{\cal SK}\otimes \cQ=\frac{SU(1,1)}{U(1)}\otimes \frac{SO(4,1)}{SO(4)}\ ,
\label{sdodi}
\ee
where a $U(1)\otimes U(1)$ isometry of $\cQ$ is gauged. In this case
the vanishing of $V$ requires a compensation of the 
$\delta\lambda ,\delta\zeta$
variations with the gravitino contribution
\be
4 k^u_\La k^v_\Si h_{uv}+ U^{\La\Si} \cP^x_\La
\cP^x_\Si=3\bar L^\La L^\Si \cP^x_\La \cP^x_\Si\ .
\label{stredi}
\ee
The moduli space of vacua satisfying (\ref{stredi}) is a four dimensional
subspace of (\ref{sdodi}).

\par
One may wonder where are the explicit mass terms for hypermultiplets. In
$N=2$ supergravity, since the hypermultiplet mass is a central charge, 
which is gauged,
such term corresponds to the gauging of a $U(1)$ charge. This is best seen
if we consider the case where no vector multiplets (and then gauginos)
are present. In this case $L^\La=L^0=1$ and the potential becomes
\be
V=4 h_{uv} k^u k^v-3 \cP^x \cP^x
\ee
where $k^u$ is the Killing vector of a $U(1)$ symmetry of $\cQ$,
 gauged by the graviphoton and  $\cP^x$ is the associated prepotential. For
$\frac{SO(4,1)}{SO(4)}$ this reproduces the Zachos model \cite{zachos}.
 The gauged $U(1)$ in this model is contained
in $SU_R(2)$ which commutes with the symmetry $SU_L(2)$ in the decomposition of
$SO(4)=SU_L(2)\otimes SU_R(2)$. This model has a local minimum at vanishing
hypermultiplet vev at which $U(1)$ is unbroken, and the extrema (at $u=1$)
(maxima) which break $U(1)$. The extremal model is when both $n_H=n_V=0$.
Still we may have a pure F-I term
\be
V=-3 \xi^2\ \ \ \xi=(\xi,0,0)
\ee
This corresponds to the gauging of a $U(1)\subset SU(2)_L$ and gravitinos
have charged coupling. This model corresponds to anti-De Sitter $N=2$
supergravity \cite{fredas}.

\section*{Acknowledgements}
A.C. and S. F. would like to thank the Institute for Theoretical Physics
at U. C. Santa Barbara for its kind hospitality and where part of this work
was completed.


\begin{thebibliography}{50}

\bibitem{SW12}
N. Seiberg and E. Witten, Nucl. Phys. B426 (1994) 19;
Nucl. Phys. B431 (1994) 484.

\bibitem{kltold} A. Klemm, W. Lerche, S. Theisen and
S. Yankielovicz, Phys. Lett. B344 (1995) 169;
P. Argyres and A. Faraggi, Phys. Rev. Lett. 74 (1995) 3931.

\bibitem{CDF}
A. Ceresole, R. D'Auria and S. Ferrara,
Phys. Lett. 339B (1994) 71, hep-th/9408036.

\bibitem{CDFVP}
A. Ceresole, R. D'Auria, S. Ferrara and A. Van Proeyen,
Nucl. Phys. B444 (1995) 92, hep-th/9502072.

\bibitem{huto1}
C. M. Hull and P.K. Townsend, Nucl. Phys. B438 (1995) 109, hep-th/9410167.

\bibitem{Wdy}
E. Witten, Nucl. Phys. B443 (1995) 85, hep-th/9503124.

\bibitem{Duff}
M. J. Duff, Nucl. Phys. B442 (1995) 47; for a recent review see also
``Electric/Magnetic Duality and its Stringy Origins'', hep-th/9509106.


\bibitem{FHSV}
S. Ferrara, J.  A.  Harvey, A.  Strominger and C.  Vafa,
 Phys. Lett. B361 (1995) 59, hep-th/9505162.

\bibitem{kava}S. Kachru and C. Vafa, Nucl. Phys. B450 (1995) 69,
hep-th/9505105.

\bibitem{cynoi} M. Bill\'o, A. Ceresole, R. D'Auria, S. Ferrara, P. Fr\`e,
 T. Regge, P. Soriani
and A. Van Proeyen, ``A Search for non-perturbative Dualities
of Local N=2 Yang-Mills Theories from Calabi-Yau Threefolds'',
preprint hep-th/9506075,
to appear on Class. and Quantum Grav.

\bibitem{KLM} A. Klemm, W. Lerche and P. Mayr, Phys. Lett. B357 (1995) 313,
hep-th/9506112.

\bibitem{bagwit} J. Bagger and E. Witten  Nucl. Phys. B222 (1983) 1.

\bibitem{hklr} N. J. Hitchin, A. Karlhede, U. Lindstrom and M. Rocek,
``HyperK\"ahler Metrics and Supersymmetry'', Commun. math. Phys. 108
(1987) 535.

\bibitem{gal}
K. Galicki, Comm. Math. Phys. 108 (1987) 117.

\bibitem{DFF}
R. D'Auria, S. Ferrara and P. Fr\'e, Nucl. Phys. B359 (1991) 705.

\bibitem{dWLVP}
B. de Wit, P. G. Lauwers and A. Van Proeyen Nucl. Phys. {B255}
(1985) 569.
\bibitem{CaDF}
L. Castellani, R. D'Auria and S. Ferrara,
Phys. Lett. 241B (1990) 57; Class. Quantum Grav. 7 (1990) 1767.

\bibitem{CFGVP} E. Cremmer, S. Ferrara, L. Girardello and
A. Van Proeyen, Nucl. Phys. {B212} (1983) 413.
\bibitem{FGP1} S. Ferrara, L. Girardello and M. Porrati, ``Minimal Higgs 
Branch for the Breaking of Half of the Supersymmetry in N=2 Supergravity'',
preprint CERN-TH/95-268, hep-th/9510074.
\bibitem{GMS}
B. Greene, D. Morrison and A. Strominger, Nucl. Phys. B451 (1995) 109,
 hep-th/9504145.
\bibitem{pols} J. Polchinski and A. Strominger, ``New vacua for Type II
String Theory'', preprint hep-th/9510227.
\bibitem{APT} I. Antoniadis, H. Partouche and T. R. Taylor,
``Spontaneous Breaking of N=2 Global Supersymmetry'', preprint hep-th/9512006.
\bibitem{lungo} L. Andrianopoli, M. Bertolini, A. Ceresole, R. D'Auria,
S. Ferrara, P. Fr\'e and T. Magri, ``N=2 Supergravity and N=2 
Yang-Mills Theory on fully General Scalar Manifolds'', in preparation.
\bibitem{CDFp} A. Ceresole, R. D'Auria and S. Ferrara, ``The Symplectic
Structure of N=2 Supergravity and its Central Extension'', Proceedings of
the ICTP Trieste Conference on ``S-Duality and Mirror Symmetry'', Miramare,
Trieste, Italy, June 1995, hep-th/9509160.
\bibitem{BDW} B. de Wit, P.G. Lauwers, R. Philippe, Su S.Q.
and A. Van Proeyen, Phys. Lett. 134B (1984) 37;
S. J. Gates, Nucl. Phys. B238 (1984) 349;
B. de Wit and A. Van Proeyen, Nucl. Phys. B245 (1984) 89.
\bibitem{dere} J. P. Derendinger, S. Ferrara and A. Masiero, Phys. Lett. 143B
(1984) 133.

\bibitem{flatpotn2}E. Cremmer, C. Kounnas, A. Van Proeyen, J.P. Derendinger,
S. Ferrara, B. De Wit and L. Girardello, Nucl. Phys. B250 (1985) 385.
\bibitem{DWNi} B. de Wit and H. Nicolai, Nucl. Phys. B208 (1982) 323.
\bibitem{CaDFb} L. Castellani, R. D'Auria and P. Fr\'e,
``Supergravity and Superstring Theory: a Geometric Perspective'',
World Scientific (1991) , Volumes 1, 2, 3.

\bibitem{fsz} S. Ferrara, J. Scherk and B. Zumino, Nucl. Phys. B121 (1977) 393.
\bibitem{ward} S. Ferrara and L. Maiani, Proc. V Silarg Symp., World Scientific
(1986); S. Cecotti, L. Girardello and M. Porrati, Nucl. Phys. B268 (1986) 295;
A. Ceresole, R. D'Auria, S. Ferrara, P. Fr\'e and E. Maina, Nucl. Phys. B268 
(1986) 317. 
\bibitem{scsc}S. Ferrara, C. Kounnas, M. Porrati and F. Zwirner, Nucl. Phys.
B318 (1989) 75.
\bibitem{ant} I. Antoniadis, Phys. Lett. B246 (1990) 377.

\bibitem{flatpotn1}E. Cremmer, C. Kounnas, S. Ferrara and D. Nanopoulos,
Phys. Lett. 133B (1983) 61.
\bibitem{zachos} C. Zachos Phys. Lett. {76B} (1978) 329.
\bibitem{fredas} D.Z. Freedman, A. Das
Nucl. Phys. {B120} (1977) 221

\end{thebibliography}
\end{document}